# Non-resonant adiabatic photon trap

S.S. Popov[1,2], M.G. Atluhanov[1], A.V. Burdakov[1,3], M.Yu. Ushkova[2]

[1] *Budker Institute of Nuclear Physics SB RAS, Novosibirsk, Russia*
[2] *Novosibirsk State University, Novosibirsk, Russia*
[3] *Novosibirsk State Technical University, Novosibirsk, Russia*

Experimental studies of the concept of an efficient radiation trap based on the adiabatic photon confinement between curved mirrors is presented in the paper. Such a trap is free from of restrictions typical to the Fabry-Perot cells, relating to the confined radiation quality, accuracy and stabilization of optical elements. Experiments have demonstrated the high efficiency of confinement, which mainly depends mainly on the reflectivity of mirrors and their surface defects on their surface.

## I. INTRODUCTION

Many photochemical and photocatalytic phenomena and their applications play an important role in the modern science and technology [1, 2]. These include the phenomenon of photoneutralization of negative hydrogen and deuterium ions, which is extremely important for thermonuclear installations [3, 4]. In practice, in use of exposure of atoms or molecules to radiation, it is either desired or required to have a high degree of saturation of absorbing transitions. This substantially affects the useful reaction rate or the yield of final product, e.g. a separated isotope. This in turn requires creation of devices to form the required radiation fluxes in isolated finite spatial zones, where one or another photoreaction occurs. It would be logical to characterize the efficiency of such devices by the ratio of confined radiant power in the reaction area to the power taken from a source, e.g. a laser. The standard way to solve the problem is creating various resonant photon traps of the Fabry-Perot cell type [5, 6, 7]. Such devices are often to be inefficient or require a very high quality of the pumping radiation and high vibration and thermal stability of the optical elements [8]. Indeed, on the one hand, the cell has to have a high quality factor which, on the other hand, hampers the radiant energy supply. It turns out that "abandonment of resonance" enables using radiation with a rather large angular spread and spectral width, injected through small holes (See II). The radiation can be confined in a given region by means of concave mirrors. In [9], a trap consisting of two spherical mirrors has been already investigated and the concept of non-resonance photon confinement was confirmed. Early, a similar approach had been used to create a folded delay line [10]. However, the system in [9, 10], firstly, has an substantially uneven intensity profile with large antinodes and dips in the confinement region, and secondly, the shape of the area is not well suited for use with particle beams or rapid gas flows.

This paper deals with an experimental study of the effectiveness of non-resonant adiabatic photon trap with an elongated area occupied by photons. This article is organized as follows. In Sec. II we give simple mathematical model for open adiabatic trap. It includes stability condition and calculation of radiation density distribution within confinement region. Section III is devoted to experiment idea and its realization for measurement of radiation storing efficiency. The Sec. IV is some discussion of obtained results. A conclusion is in section V.

## II. MATHEMATICAL MODEL OF OPEN ADIABATIC TRAP

### A. Simplest two-dimensional geometry and confinement condition

Let us consider the case of two-dimensional motion of a photon or a beam between the concave and flat mirrors and a flat one (see Fig. 1). As can be seen from the Fig.1a, with each reflection from the upper mirror, the photon gets a horizontal momentum difference towards the side where the distance $F$ to the lower mirror is larger. Under conditions of small deviation of the photon motion direction from the vertical, it will tend to the central "equilibrium" position. Let us find out the law of the photon motion along the horizontal in this optical trap to determine the confinement boundaries and stability conditions. Let us define the photon position immediately after the nth reflection with the abscissa of the reflection point $x_n$, its height $F(x_n)$ and the angle between the vertical and the photon speed $\beta_n$ (see Fig. 1a). Then the horizontal motion is described by the following system of equations:

$$x_{n+1} - x_n = \left(F(x_{n+1}) + F(x_n)\right) tg \beta_n. \qquad (1)$$

$$\beta_{n+1} - \beta_n = 2 arctg\left(\frac{dF(x_{n+1})}{dx}\right). \qquad (2)$$

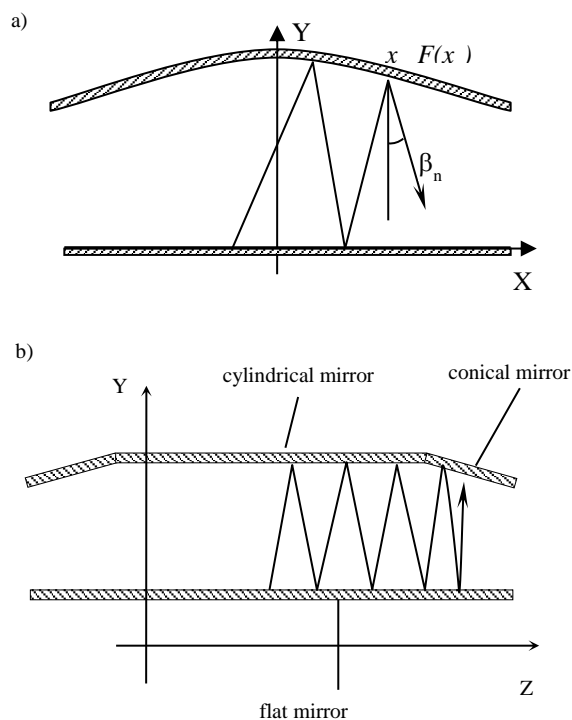

FIG. 1. Basic optical scheme of elongated adiabatic photon trap: a) transverse view, b) longitudinal view.

Let us assume that $\left|\frac{dF(x_{n+1})}{dx}\right| \ll 1$, then a combination of equations (1) and (2) enables separation of a discrete integral of motion,

$$\sum_n tg\beta_n(\beta_{n+1} - \beta_n) = \sum_n \frac{2(x_{n+1} - x_n)}{F(x_{n+1}) + F(x_n)} \frac{dF(x_{n+1})}{dx}. \qquad (3)$$

In the case of sufficient smoothness of the upper mirror and small steps, such that

$$|\Delta F| \ll F, \quad \left|\frac{dF}{dx}\right| \ll 1, \quad |\Delta \beta| \ll 1, \qquad (4)$$

the integral sums in (3) are approximately transformed into $\ln\frac{\cos\beta_0}{\cos\beta} = \ln\frac{F(x)}{F(x_0)}$ or into a standard adiabatic invariant

$$F(x)\cos(\beta) = const. \qquad (5)$$

The latter defines the area occupied by the photons provided that the amplitude of the quasi-periodic motion of photons is not increasing.

To investigate the stability we linearize the system of (1) and (2) and obtain

$$x_{n+1} - x_n = 2F(0)\beta_n \qquad (6)$$

$$\beta_{n+1} - \beta_n = 2\frac{d^2F(0)}{dx^2}x_{n+1}. \qquad (7)$$

Substituting one equation to the other, we obtain the following linear recurrence:

$$x_{n+2} - 2x_{n+1} + x_n = 4F(0)\frac{d^2F(0)}{dx^2}x_{n+1} = -4F(0)\frac{x_{n+1}}{r}, \qquad (8)$$

where $r$ is the radius of curvature of the upper mirror. Equation (8) is a difference scheme for an oscillation system with a unit time step and a eigenfrequency $\omega_0 = 2\sqrt{\frac{F(0)}{r}}$. Obviously, the solution can be represented as

$$x_n = A \cdot q^n, \qquad (9)$$

where $q$ is a complex value. Then for $q$ we have

$$q_{1,2} = 1 - \frac{2F(0)}{r} \pm \sqrt{\left(1 - \frac{2F(0)}{r}\right)^2 - 1}. \qquad (10)$$

The difference scheme is stable if $|q| \le 1$, whence, given the non-negativity of $\frac{F(0)}{r}$, we obtain the condition of "geometric" photon confinement

$$F(0) < r, \quad \omega_0^2 < 4. \qquad (11)$$

The above restriction coincides with the classical condition for stability of a cavity formed by a concave spherical mirror and a flat one, obtained via consideration of an infinite-degree composition of the matrices of ideally focusing mirrors (e.g. see [11]). However, the recurrent approach allows us to evaluate the effect of the non-linearity. Indeed, with sufficiently large angles of the beam tilt, when the approximation $tg\beta_n \approx \beta_n$ is violated significantly, but an difference $|\beta_{n+1} - \beta_n|$ is still much smaller than the amplitude of the angle variation of the angle, expression (8) takes the following form:

$$x_{n+2} - 2x_{n+1} + x_n = -4F(0)\frac{x_{n+1}}{\cos^2(\beta_{n+1})r}. \qquad (12)$$

The cosine in the right-hand side provides a local non-linear increase in frequency. As can be seen from Fig. 1a, the maximum frequency occurs near its equilibrium, where photons have the maximum angle to the vertical. Then the stability condition has to be strengthened.

$$F(0) < r \cdot (\cos\beta_{max})^2. \qquad (13)$$

This correction can be substantial near the stability boundary.

As one can see, the radius of the curvature of the mirrors is crucial for the photon confinement and thus it is impossible to use solely flat elements mated with a finite angle. Such a case is considered in [12]. In a practically important three-dimensional case with satisfied (4) and (13), the photon movements or beam evolutions along the axes X and Z can obviously be considered independently from each other. The curvatures of the upper mirror along these lines can be significantly different. It would be somewhat more complicated to calculate the case with two non-planar mirrors, but we will not give details here.

The above estimates are sufficient for conceptual development of an efficient photon trap for neutralization of negative ion flow or photochemical applications in gaseous media, which slightly interact with radiation. A possible geometry for such applications is shown in Fig. 1. The trap consists of an upper cylindrical mirror mated at the ends with, for example, conical mirrors. The radii of the cylinder and the larger bases of the cones are equal. For the sake of simplicity, the lower mirror may be flat. The end mirrors can be of non-conical shape. They are to provide slow reduction in the distance between the upper and lower mirrors when moving from the center of the trap. If the upper mirrors are reflected relatively to the lower one, we get an equivalent system of identical non-planar mirrors. Movement of photons in such a system along the Z axis is substantially anharmonic, and hence no practically parasitic intensity peaks can arise. In other words, when the beams move quasi-periodically along the Z axis, they experience some mixing. An example of numerical modeling of such systems see in III.B Note that the longitudinal movement can be unstable. However, if the length of the trap is large enough, the time for a photon to leave the trap will exceed the inverse decrement of the radiation because of absorption in repeated reflections from the mirrors. If necessary, the stability in the longitudinal direction can be improved using blending of the central mirror with the end ones. In this case, the latter can be, for example, spherical or toroidal. Further, we assume the influence of photon ensemble instability along the Z axis to be unimportant in comparison with absorption in reflections.

## B. Distribution of radiation density in the trap

In the below approach to the experimental validation of the concept, it is important to find the dependence of the local concentration of photons inside the trap on the reflectivity.

Let the radiation be injected locally with a small angular spread, a positive initial velocity and near the left turning point. Then we have a continuity equation with a dissipation of the particles in the stream:

$$\langle v_z(z)\rangle n(z) = n_0 v_0 \frac{R^{m(z)}+R^{N-m(z)}}{1-R^N}. \quad (14)$$

Equation (14) is the geometric progression sum. Here the zero subscript corresponds to the point of injection; $m(z)$ is the number of reflections experienced by photons when moving from the source towards the vicinity of the point z; $\langle v_z(z)\rangle \approx \left\langle \sqrt{c^2-v_y^2-v_x^2} \right\rangle$ is the speed longitudinal projection averaged over the transverse motion period; N is the number of reflections over the full period. The dissipation produced by non ideal reflections. The second term in the numerator is associated with returning to the area of consideration after the first turning point.

In the case of multiple injection points, we have a set of functions $m_i = m(z,z_{0i},v_{0i})$; $N_i = N(z_{0i},v_{0i})$ and the following distribution of the photon density:

$$n(z) = \sum_i \frac{n_{0i}v_{0i}}{v_{zi}(z)} \frac{R^{m_i(z)}+R^{N_i-m_i(z)}}{1-R^{N_i}}. \quad (15)$$

One can calculate the explicit form of the functions $m_i(z,z_{0i},v_{0i})$; $N_i(z_{0i},v_{0i})$; $\langle v_z(z)\rangle$ taking into account the adiabaticity and quasi-periodicity of the motion, but in this study only their existence essential. In a case of practical importance for us, when $N_i \gg 1$, but $1-R^{N_i} \ll 1$, from (15) we obtain a simple uniform dependence of the photon local density on the reflectivity of the mirrors,

$$n(z) = \sum_i \frac{n_{0i}v_{0i}}{v_{zi}(z)} \frac{2-N_i(1-R)+O\left(((1-R)N_i)^2\right)}{N_i(1-R)-N_i(N_i-1)(1-R)^2/2+O\left(((1-R)N_i)^3\right)}$$
$$\approx \frac{2}{(1-R)} \sum_i \frac{n_{0i}v_{0i}}{v_{zi}(z)N_i}\left(1+O\left(((1-R)N_i)^2\right)\right) \quad (16)$$

The $O(\xi^n)$ value is such that $\lim_{\xi\to 0}\frac{O(\xi^n)}{\xi^n}<\infty$. Note that relationship (16) also holds in the case of a random arrangement of injection centers for all points to the right of them, provided an efficient source is introduced at the left turning point with a renormalization of the current density $n_{0i}v_{0i}\to n_{0i}v_{0i}R^{-k_i}$, where $k_i$ is the number of reflections between the left stop point and the actual source.

## III. Experimental study of efficiency of photon confinement in open adiabatic trap

### A. Mirrors and general idea of the experiment

For the experimental study, a set of mirror elements on substrates of single-crystal silicon with a multilayer dielectric coating was made. It included cylindrical segments and a few spherical ones. The length of one segment was 50 mm; the width of the reflecting surface was about 30 mm. The total size of the mirror could be up to 250 mm with a confinement zone of 150 ÷ 200 mm. The rated reflectivity of some mirrors was 0.999. One version of the arrangement of the big mirror is shown in FIG.2.

In the case of stable confinement, the efficiency as the ratio of the confined power to the injected one is obviously defined by the value $1/(1-R)$. In this study, to measure it we investigated how the radiation decay dynamics in the trap with fast switch-off of the pumping depended on various additional attenuators placed between the mirrors. We also controlled the steady-state level of accumulated radiation at constant pumping. Assuming that the steady-state level obeys equation (16), one can determine the value $1-R$ (see (17)). The simple and obvious photometric approach applied by the authors in [9], basing on the measurement of the parasitic scattering from the surface of the mirrors, turned out to be inconvenient for this system for two reasons. The first is the large number of joints, which give a strong stray light, and the second is the large area of mirror, which complicates the visual-field calibration of photographic

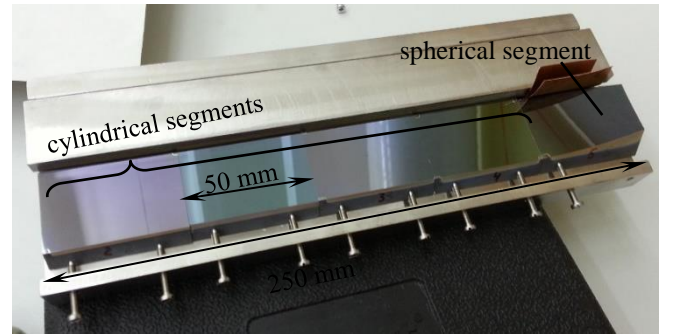

FIG. 2. One of possible versions of mating the mirror elements.

sensitivity and consideration of the background radiation.

## B. Experimental setup

The experiment setup is shown in Fig.3.

The first six elements provide fast radiation modulation. Special attenuator 2 forms a narrow beam of photons from the radiation emerging from the laser head 1. Then the beam crosses glass beam splitter 3, rotated through the Brewster angle to the direction of the motion of photons. Next in the way is Pockels cell 4 with polarizer 5. The radiation reflected back by mirror 6 crosses polarizer 5 for the second time. Thus, the radiation returned to the cell is highly polarized, and with no voltage on the cell, the radiation from beam

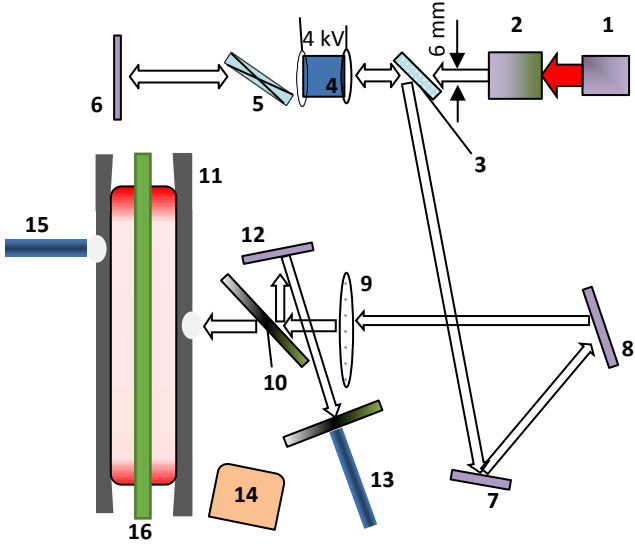

FIG. 3. Scheme of measurement of efficiency of adiabatic trap. 1) laser head; 2) collimator; 3) beam splitter; 4) Pockels cell; 5) polarizer; 6,7,8,12) mirrors; 9) lens; 10,12) light filters; 11) photon trap; 13) fiber for monitoring input radiation; 14) CCD camera; 15) signal fiber, 16) attenuator

splitter 3 in the direction of mirror 7 is not reflected. With a voltage of about 4 kV, the cell converts the linear polarization to the circular one, and about 10% of the power gets to mirror 7. When the cell voltage is switched off, such a shutter enables a rather fast blocking of the radiation flux on mirror 7 and further on. The double passage of polarizer 5 ensures a higher degree of polarization and thus the better contrast of the radiation flux entering the trap between the closed and open states of the shutter. Mirrors 7 and 8 and lens 9 direct the radiation through the infrared light filter and beam splitter 10 to the inlet of about 300 microns in diameter in trap 11. Mirror 12 directs the radiation part reflected by the beam splitter into optical fiber 13 for monitoring of the input radiation. The Digital Camera SDU285 [13], recording the image of the inlet, made it possible to monitor the radiation getting into it. Optic fiber 15 at the other opening collects a small fraction of the stored radiation for detection and analysis. Avalanche photodiodes with a time resolution of about 2 ns recorded the dynamics of the input and confined radiation. Arrays of tungsten wires with a diameter $d$ = 15 μm were used as additional attenuators. Each attenuator is a frame with vertically tensioned wires with a certain frequency.

The attenuation coefficient was assumed to be $\alpha=\sigma d$, where $\sigma$ is the frequency of the wires per unit length. The radiation source was an industrial ytterbium fiber laser [14] with a wavelength of 1070 nm and a spectral width of about 7 nm.

An example of numerical simulation of intensity distribution in the middle plane between mirrors obtained by ZEMAX code is shown in Fig.4. Here radiation beam with an angular spread about 3 degree is injected in YZ-plane (See Fig.1) at an angle 94 degree relative to the Z-axis. The distance between mirrors is 12 cm. The total power is 1 W, reflectance 0.995. As seen, the intensity can be inhomogeneous along the trap. Due to the inclined injection and not very high reflectance, the number of photons with negative z-coordinate exceeds the number of those with the negative one.

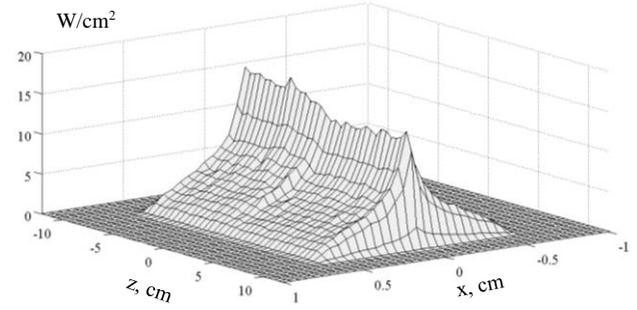

FIG. 4. Numerical calculation for intensity in the middle plane between mirrors. Angular spread is about 3 degree, photon beam is tilted in YZ -plane about 94 degree (See Fig.1).

## C. Measuring the confinement efficiency

The waveforms of exit radiation out of the trap through a small aperture for one series of experiments are shown in Fig. 5. In these experiments, the distance between mirrors was about 12 cm; the beam angle to the axis Z was about 85 degree. The confinement area was covered by five segments on each mirror surface with four joints. In Fig.5a the set of thin solid lines correspond to the dynamics of the radiation in the outlet of the trap with different values of additional attenuation α. A set of waveforms in Fig.5b shows the dynamics of the shutting of the input radiation. In the time interval from 0 to 100 ns, one can see the steady-state level of the trap pumping. The shutting lasts for about 20 ÷ 30 ns. Then an exponential decay of the radiation level began, shown by the approximating thick curves. They agree quite well with the waveforms. The non-zero asymptotic level at times of the order of 1 µs is caused by the incomplete contrast of the input radiation switch. As one can see, there is a significant dependence of the decrement and steady-state levels on the additional attenuation introduced into the trap. The on-mirror losses proper can be estimated as $10^{-3}$. A more accurate measurement of the efficiency can be done with consideration of the steady-state radiation levels taken from (16) with various losses per pass of $1-R+\alpha$. Finally we obtain

$$1-R = \frac{n(z,\alpha_k)\alpha_k - n(z,\alpha_l)n\alpha_l}{n(z,\alpha_l)-n(z,\alpha_k)}. \quad (17)$$

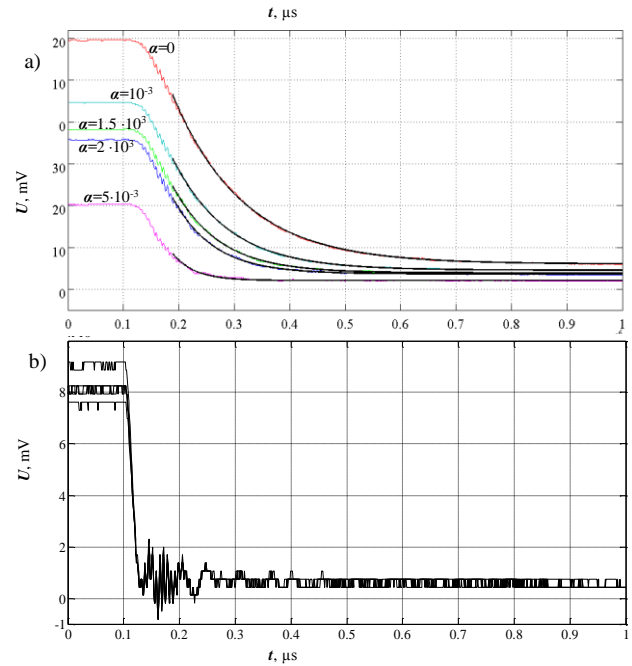

FIG. 5. Typical waveforms : a) confined radiation with different additional attenuation α (colored lines), thick black lines - decaying exponential curves fitting waveforms; b) set of corresponding input radiation waveforms from monitor fiber 13 (See fig. 3);.

Alternatively, the confinement efficiency can be extracted from the temporal dynamics of radiation decay after injection termination. Indeed, we obtain from (16) with time scales greater than the period of the longitudinal motion $T$

$$n(z,t) \approx \frac{2}{(1-R)} \sum_i \frac{n_{0i} v_{0i} R^{N_i \frac{t}{T_i}}}{v_{zi}(z) N_i}$$
$$\approx \frac{2}{(1-R)} \sum_i \frac{n_{0i} v_{0i} \exp\left(N_i \frac{t}{T_i}(R-1)\right)}{v_{zi}(z) N_i} \quad (18)$$

In our case, with injection of a beam with a small angular spread, the average frequency of the reflections $\frac{N_i}{T_i}$ is almost the same for all injected beams. Then from (18) we have

$$n(z,t) \propto \exp\left(\left\langle\frac{N_i}{T_i}\right\rangle t(R-1)\right), \quad (19)$$

where $\left\langle\frac{N_i}{T_i}\right\rangle$ is a certain characteristic frequency of reflections for all photons. Thus, comparing the exponential decrements obtained from (19) with various attenuators, one can independently calculate the efficiency of radiation confined in the trap:

$$1-R = \frac{\tau_l \alpha_l - \tau_k \alpha_k}{\tau_k - \tau_l}. \quad (20)$$

When waveforms with and without various additional attenuators were compared, both the measurement methods gave consistent results: $1-R = (2.3 \pm 0.5) \cdot 10^{-3}$.

## IV. DISCUSSION

The good agreement of the steady-state level and exponential decay in Fig. 5 confirms the validity of (20) as well as a possibility of the adiabatic confinement in the open mirror system. Note that exponential temporal decaying is obvious for the case Fabry-Perot accumulation (See for example [15]). Here it required additional justification, because we observe radiation out of the local point of region occupied by radiation in contrast to [15].

Significant changes in the mutual position of the mirrors, a few centimeters in the distance and a few degrees in the rotation, did not change the confinement efficiency. This shows the insensitivity of the trap operating to both the radiation quality and the accuracy of adjustment of the mirrors. Using ratio for the sum of geometric sequence first $j$ terms $(1-R^j)/(1-R)$, we obtain that about 90% of accumulated radiation power contains in the first 1000 reflections. It shows a significant attractiveness of such an approach. For comparison, in project based on Fabry-Perot cell [6] the effective reflections amount is about 500 at assumed reflectivity of mirrors 0.9996 and maximum mirror misalignment ~20 µrad.

The main contribution to the error of storing efficiency measurement is apparently given by the sensitivity drift of photodiodes, especially of the monitor channel. In addition, there may be a random effect of attenuators in the first radiation passes in the trap, when the beam has not yet expanded sufficiently. In this case, the beam attenuation on these passages may vary slightly. The number of such passes depends on the distance where the beam width exceeds the spacing between adjacent wires. In this case, this value is about 10 passes. Accidental losses on the dust suspended between the mirrors may also be substantial. No special acts for dust filtering have been taken.

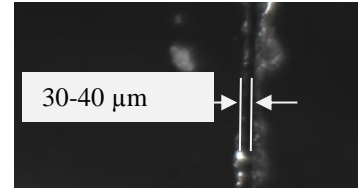

FIG. 6. Defects of mirror surface at joint of two elements.

It is obvious that joints between the mirror elements have a negative effect on the radiation confinement. An example of joint is shown in Fig. 6. The period-averaged probability that a photon gets on a joint and is absorbed can be estimated as

$$\mu_{sl} \sim \frac{S_{sl}}{\langle x_{n+1} - x_n \rangle} k, \quad (21)$$

where $\langle x_{n+1} - x_n \rangle$ is a characteristic interval of impact parameters; $S_{sl}$ is the thickness of the joint; $k$ is the amount of joints the photon encountered over the period. Let us compare it with the absorption on the mirrors in a period, $\mu \approx R^N$, given that $\langle x_{n+1} - x_n \rangle \sim 2L/N$ ($L$ is the length of the confinement area):

$$\frac{\mu_{sl}}{\mu} \sim \frac{S_{sl} k N}{2L(1-R^N)} \approx \frac{S_{sl} k}{2L(1-R)}. \quad (22)$$

Let the total number of joints be 8, and $S_{sl} \approx 40$ um, and we then obtain $\frac{\mu_{sl}}{\mu} \sim 1$. Thus, the absorption on the junctions can be significant. Consequently, to make the confinement more efficient, one needs both to improve the reflectivity of the mirrors and to reduce the density of the joints or their width.

## V. CONCLUSION

Traditionally, the systems based on the Fabry-Perot cavities and proposed for concentrating radiant energy have a number of strong limitations. The record parameters obtained in [15, 16] are due to the outstanding characteristics of the mirrors, small spatial scales, and low energy density. Advances in the creation of sufficiently effective

Fabry-Perot cells in gravitational interferometers are based, among other things, on a very complicated system of mounting and stabilization of optical elements [17, 18].The adiabatic trap we suggest in this paper is obviously insensitive to the most intrinsic limitations inherent to cavities. This trap enables research when the high localized density of radiant flux is required and the radiation quality does not matter. This may be the multiphoton photochemistry [2], isotope separation, and others. Such a trap is particularly attractive for photoneutralization of beams of negative ions in fusion applications [3]. Recent sufficiently successful experiments on photon neutralization of negative ion beam have been carried out [19]. In this work the adiabatic photon trap from the same mirrors and laser were used.

The study was funded by a Russian Science Foundation grant (project No. 14-50-00080)

**REFERENCES**

1. E.V. Shpolsky "Modern photochemistry" Advances in Physical Sciences 36 (4) 295-310 (1993).

2. G.N. Makarov "Low-energy methods of molecular laser isotope separation" Advances in Physical Sciences 58 670-700 (2015)

3. J.H. Fink, A.M. Frank. "Photodetachment of electrons from negative ions in a 200 keV deuterium beam source", Lawrence Livermore Natl. Lab., UCRL-16844, (1975).

4. G. Abdrashitov, Yu. Belchenko, A. Ivanov, S. Konstantinov, A. Sanin, I. Shikhovtsev, O. Sotnikov, and N. Stupishin Emission properties of inductively driven negative ion source for NBI // AIP Conference Proceedings 1771, 030013 (2016); doi: 10.1063/1.4964169

5. V. Vanek, T. Hursman, D. Copeland, D. Goebel, "Technology of a laser resonator for the photodetachment neutralizer", Proc. 3rd Int. Symposium on Production and Neutralization of Negative Ions and Beams, (Brookhaven, 1983). P.568-584.

6. M. Kovari, B. Crowley. "Laser photodetachment neutraliser for negative ion beams",Fusion Eng. Des., 85, P.745–751. (2010)

7. W. Chaibi, C. Blondel, L. Cabaret, C. Delsart, C. Drag, A. Simonin, "Photoneutralization of negative ion beam for future fusion reactor", Negative Ions Beams and Sources: 1st International Symposium, AIP Conference Proceedings. V. 1097. P. 385 (2009).

8. D. Fiorucci , W. Chaibi , C. N. Man and A. Simonin. "Optical Cavity Design for application in NBI systems of the future generation of Nuclear Fusion reactors". Programme and Book of Abstracts of 4th International Symposium on Negative Ions, Beams and Sources Garching, Germany 6-10 October 2014. P. 05-04 (http://www.ipp.mpg.de/3768738/programme_books_of_abstracts.pdf).

9. S.S. Popov, M.G. Atluhanov, A.V. Burdakov, M.Yu. Ushkova. "Experimental study of non-resonant photon confinement in a system of spherical mirrors". Optics and Spectroscopy. 121, p. 160-163 (2016).

10. Donald R. Herriott and Harry J. Schulte, "Folded Optical Delay Lines," Appl. Opt. 4, 883-889 (1965).

11. H. Kogelnik and T. Li, "Laser Beams and Resonators," Appl. Opt. 5, 1550-1567 (1966).

12. I. A. Kotelnikov, S. S. Popov, and M. Romé. "Photon neutralizer as an example of an open billiard", Phys. Rev. E **87**, 013111, (2013).

13.SpecTeleTehnika.: http://www.sptt.ru/sptt/catalog.php?mod=sdu285

14. http://www.ntoire-polus.ru/HP%20fiber%20laser.pdf

15. G. Rempe, R. J. Thompson, H. J. Kimble, and R. Lalezari. "Measurement of ultralow losses in an optical interferometer", Optics Letters, Vol. 17, Issue 5, pp. 363-365 (1992).

16. Christina J. Hood, H. J. Kimble, and Jun Y. "Characterization of high-finesse mirrors: Loss, phase shifts, and mode structure in an optical cavity", Phys. Rev. A 64, 033804 (2001).

17. J Aasi, B P Abbott, R Abbott, and LIGO team. "Advanced LIGO", Classical and Quantum Gravity 32, 074001 (2015).

18. F Acernese, M Agathos, K Agatsuma, and VIRGO team "Advanced Virgo: a second-generation interferometric gravitational wave detector", Classical and Quantum Gravity 32, 024001 (2015).

19. M. G. Atlukhanov, A. V. Burdakov, A. A. Ivanov, A. A. Kasatov, A. V. Kolmogorov, S. S. Popov, , M. Yu. Ushkova, and R. V. Vakhrushev. AIP Conference Proceedings 1771, 030024 (2016); doi: http://dx.doi.org/10.1063/1.4964180